*Article*

# An Innovative Approach to Addressing Childhood Obesity: A Knowledge-Based Infrastructure for Supporting Multi-Stakeholder Partnership Decision-Making in Quebec, Canada

**Nii Antiaye Addy [1,*], Arash Shaban-Nejad [2,3], David L. Buckeridge [3] and Laurette Dubé [1]**

[1] McGill Center for the Convergence in Health and Economics (MCCHE), Desautels Faculty of Management, McGill University, Montreal, Quebec H3A 1G5, Canada;
E-Mail: laurette.dube@mcgill.ca

[2] School of Public Health, University of California, Berkeley, CA 94720, USA;
E-Mail: arash.shaban-nejad@mcgill.ca

[3] Department of Epidemiology and Biostatistics, McGill University, Montreal, Quebec H3A 1A3, Canada; E-Mail: david.buckeridge@mcgill.ca

* Author to whom correspondence should be addressed; E-Mail: nii.addy@mcgill.ca;
Tel.: +1-514-398-3299; Fax: +1-514-398-8963.

**Abstract:** Multi-stakeholder partnerships (MSPs) have become a widespread means for deploying policies in a whole of society strategy to address the complex problem of childhood obesity. However, decision-making in MSPs is fraught with challenges, as decision-makers are faced with complexity, and have to reconcile disparate conceptualizations of knowledge across multiple sectors with diverse sets of indicators and data. These challenges can be addressed by supporting MSPs with innovative tools for obtaining, organizing and using data to inform decision-making. The purpose of this paper is to describe and analyze the development of a knowledge-based infrastructure to support MSP decision-making processes. The paper emerged from a study to define specifications for a knowledge-based infrastructure to provide decision support for community-level MSPs in the Canadian province of Quebec. As part of the study, a process assessment was conducted to understand the needs of communities as they collect, organize, and analyze data to make decisions about their priorities. The result of this process is a "portrait", which is an epidemiological profile of health and nutrition in their community. Portraits inform strategic planning and development of interventions, and are used to assess the impact of interventions. Our key findings indicate ambiguities and disagreement among MSP decision-makers regarding causal relationships between actions and outcomes, and the relevant data needed for making decisions. MSP decision-makers expressed a desire for easy-to-use tools that facilitate the collection, organization, synthesis, and analysis of data, to enable decision-making in a timely manner. Findings inform conceptual modeling and ontological analysis to capture the domain knowledge and specify relationships between actions and outcomes. This modeling and analysis provide the foundation for an ontology, encoded using OWL 2 Web Ontology Language. The ontology is developed to provide semantic support for the MSP process, defining objectives, strategies, actions, indicators, and data sources. In the future, software interacting with the ontology can facilitate interactive browsing by decision-makers in the MSP in the form of concepts, instances, relationships, and axioms. Our ontology also facilitates the integration and interpretation of community data, and can help in managing semantic interoperability between different knowledge sources. Future work will focus on defining specifications for the development of a database of indicators and an information system to help decision-makers to view, analyze and organize indicators for their community. This work should improve MSP decision-making in the development of interventions to address childhood obesity.

## 1. Introduction

In addressing the complex problem of childhood obesity, one widespread means for deploying policies in a whole-of-society strategy is through multi-stakeholder partnerships (MSPs) [1], defined here as "voluntary cooperative arrangements between actors from the public, business and civil society that display minimal degree of institutionalization, have common nonhierarchical decision-making structures and address public policy issues" [2]. Assuming an instrumental perspective of stakeholder theory in which stakeholders constitute organizational life [3–5], some MSP proponents note that flexible governance structures and diverse expertise of multiple actors contribute to making such partnerships more effective in realizing desired policy outcomes, as compared with traditional single-actor structures [6]. According to such viewpoints, and of particular interest in this paper, the knowledge that diverse stakeholders bring to decision-making in MSPs is critical to their success in tackling childhood obesity.

However, decision-making in MSPs is fraught with challenges, as stakeholders are faced with complexity, and have to deal with disparate conceptualizations of knowledge across multiple sectors and levels, and diverse sets of indicators and data, to name a few [7]. For example, in the Canadian province of Quebec, as is the case in other parts of the world, MSPs that develop and implement interventions to promote healthy lifestyles are faced with challenges of using knowledge and data within complex systems that span health and non-health domains for decision-making and governance [8]. Despite the challenges, improving the functioning of such MSPs is vital to bringing about changes to health-related behaviors. Changes in such behaviors, including physical activity and healthy nutrition, are necessary to slow the "obesity epidemic", and could prevent up to 80% of heart disease, diabetes and respiratory diseases and 40% of cancers [9].

The opportunities presented and challenges faced in using multi-stakeholder partnerships (MSPs) to address the complex issue of childhood obesity is exemplified by the case of Quebec. In 2002, within the framework provided by Quebec's *Governmental Action Plan (GAP) on lifestyles and weight-related problems* [8], the government of Quebec established a partnership with Canada's largest private family foundation, the Lucie and André Chagnon Foundation [10], to invest a total of 480 million Canadian dollars from 2007 to 2017 into obesity prevention. A nongovernmental organization, Québec en Forme (QEF) [11,12], was established as a mechanism for channeling matched funds from the government and the Chagnon Foundation to support community level actors to conceive, develop and implement activities for healthy living among children and youth. Specifically, QEF's mission is to improve physical activity and healthy eating among children and youth (aged 0-17) in Québec by supporting "local partner groups" or *regroupements locaux de partenaires (RLPs)*—which are geographically defined community networks—as they develop three-year strategic plans (*i.e.*, 2014–2017), and annual action plans with interventions to improve physical activity and healthy eating in their own communities [13]. There has been expansion of local community level partnerships in Québec en Forme, with expansion also at regional and provincial levels, as shown by summarized statistics for the period 2007–2013 (Table 1).

**Table 1.** Québec en forme (QeF) Partnerships, 2007–2013 *.

| Number | 2007 | 2008 | 2009 | 2010 | 2011 | 2012 | 2013 |
|---|---|---|---|---|---|---|---|
| Local partners groups (*regroupements locaux de partenaires, RLPs*) | 35 | 35 | 71 | 110 | 140 | 152 | 157 |
| Administrative regions | 8 | 8 | 16 | 17 | 17 | 17 | 17 |
| Regional projects | NA | NA | NA | NA | 19 | 25 | 25 |
| Province-wide projects | NA | NA | NA | NA | 18 | 19 | 36 |
| Amount invested (Millions of Canadian Dollars) | 5.7 | 4.5 ** | 8.1 | 15.5 | 16.6 | NA | NA |

Notes: * See http://www.quebecenforme.org/en/about-us/history/quebec-en-forme-s-expansion-by-the-numbers.aspx.
Data as of 31 August for 2007; 31 March in each given year for 2008–2010; 30 June each year for 2011 and 2012;
and 30 May for 2013; ** Over seven-month period.

From an instrumental stakeholder perspective [3–5], this paper reports on a study to define specifications for a knowledge-based infrastructure that will provide decision support for community-level MSPs in Quebec, as they work towards a shared goal of facilitating changes in physical activity and food consumption behavior. As part of the study, a process assessment was conducted to understand the needs of communities as they collect, organize, and analyze data during their development of "*portraits*", which are epidemiological profiles of the health and nutrition needs of a neighborhood. The portraits inform stakeholders' decision-making about their priorities, strategic planning, and the development and evaluation of interventions. The purpose of this paper is to describe and analyze the development of a knowledge-based infrastructure to support MSP decision-making processes. In the next section of this paper, we present the context of the processes studied in Quebec, and we then present the potential value of conceptual mapping and ontological development. We then describe our study approach, present our findings about the needs of the MSPs that can be supported and addressed by a knowledge-based infrastructure, and report some initial mappings of processes, and concepts. We conclude by outlining next steps for our study.

## 2. The Quebec Context and the Value of Conceptual Mapping and Ontology Development

In Quebec, as elsewhere, the diverse organizational and individual actors in community networks engage in complex and dynamic processes. At the organizational level, each community network consists of organizations from multiple domains [14], including early childhood; primary and secondary school board and schools; municipal organizations, cities, and boroughs; community organizations, including those concerned with sports/leisure, food, youth, childhood, and families; centers for health and social services; and miscellaneous organizations concerned with food, *etc.* At the individual level, various stakeholders play an advisory role in community networks, in the choice of information and data considered for developing and analyzing community epidemiological profiles, and in prioritizing objectives, strategies and actions to focus on during planning and evaluation.

As part of their mobilization, the diverse actors in community networks engage in planning processes to request funding from QEF. An initial stage of community network planning processes is to create an epidemiological profile of the community in terms of ongoing activities, and environmental/contextual factors that may influence physical activity or healthy eating Creating the epidemiological profile requires analysis and organization of a considerable amount of data and information, and is currently performed by community network members, drawing mostly from studies/reports, including those provided by QEF or identified by individual members.

The creation of an epidemiological profile is a difficult task for these community networks, however, whose members may have little experience performing this sort of task. Yet, the development of the profiles is critical for informing the MSPs' strategic plans and interventions to prevent childhood obesity. Given challenges that they face in developing epidemiological profiles, the following question arose: *How could QEF support decision-makers in community networks to access and analyze relevant data for developing their epidemiological profiles?* To address the question, QEF commissioned the study that informed this article, with the goal of finding innovative approaches to supporting decision-makers in the community networks as they develop their community epidemiological profiles. We particularly note that the diversity of knowledge domains that inform community networks poses challenges, in that there may be ambiguities of definitions, understandings, and data sources used for decision-making. As we discuss further below, a knowledge-based system can support the MSP process, but can only help if it contains a semantic framework for a consensus knowledge that can bridges disparate conceptual models and data sources from various interrelated domains. We present conceptual mapping and ontology development as innovative solutions that contribute to community network's planning processes as they request funding from QEF. For this study, a deeper understanding of community processes went hand in hand with conceptual mapping and ontology development.

A concept map provides a simple intuitive medium for capturing knowledge by representing data and information in a domain through diagrams consisting of concepts, which it connects with labeled arrows in a hierarchical (taxonomical) structure. The relationship between concepts can be articulated in linking phrases such as "influences", "causes", "requires", "contributes to", or "has effect on" [15]. Ontologies—defined as a formal explicit specification of shared conceptualization [16]—provide a semantic framework for knowledge dissemination, exchange, and discovery via reasoning and inferencing. Ontologies capture the knowledge in a domain of interest through concepts, instances and the relationships (taxonomic and associative). The taxonomic relationships organize concepts into sub/super (narrower/broader) concept tree structure, while associative relationships relate instances of defined concepts. Enriching the conceptual model into ontologies using a formal knowledge representation language (e.g., OWL 2 Web Ontology Language [17]) supports knowledge integration necessary for community-level decision-making, and there are existing efforts at using such innovations to address policy challenges.

## 3. Existing Methodologies

Childhood obesity is a multifaceted health issue, which demands an integrated transdisciplinary approach for analyzing different risk factors, indicators (e.g., population health, socio-economic and political indicators), interventions, barriers, opportunities, levers and resistances. Given the wide array of data and information sources accessible from multiple settings, our aim is to design and develop an integrated knowledge-based system to support semantic analytics for timely decision making to fight childhood obesity. Most of the current research projects (e.g., EPODE [18], SPLASH [19]) focusing on automatic surveillance of obesity and its associated disorders rely heavily on databases and syntactic approaches. Also there are other efforts [20–23] on using ontologies and knowledge-bases to provide semantic support for computerized systems for childhood obesity prevention and healthy living. Each of these projects attempts to cover certain factors and parameters in the domain, with different level of formalisms and different degree of granularities. There are also various approaches [24,25] on incorporating data and knowledge integration tools and techniques to generate a consistent inter-linked view of a domain. One distinguished feature of our research in comparison with the existing models is our emphasis on the process maps capturing the necessary knowledge on opportunities, levers and resistance faced in the community for influencing healthy eating and physical activity.

We are also reusing some of the existing controlled vocabularies (e.g., AGROVOC (FAO agricultural thesaurus [26]) and ontologies (e.g., COPE [20], and SNOMED CT [27]) under the compliance with Open Biological and Biomedical (OBO) [28] standards and guidelines.

## 4. Method

Given the context presented, this study combined two methodological approaches (Table 2): (i) assessment of MSP decision-making processes to understand their needs during such processes [29]; and (ii) conceptual mapping and ontological modeling to formally represent the core knowledge in the MSP process in a format that could be used to support a knowledge-based system for integrating and interpreting community data. The study was conducted during the period September 2013 to November 2014. First, we obtained data from a convenience sample of 3 QEF-funded MSPs in Montreal, to understand how they developed epidemiological profiles to inform their 2014–2017 strategic planning. Drawing upon interviews, participant-observer techniques, and focus group discussions with the MSP members and QEF staff, we sought to identify the key stakeholders (*i.e.*, the multiple community organizations and QEF actors implicated in the development of each community epidemiological profile), the factors shaping their decision-making processes during the creation of the profiles, to assess the potential role of a knowledge-based infrastructure to support the processes. We explored their information needs, including the questions they were seeking to answer in developing epidemiological profiles, the data sources and information they had available, and how the data sources and information are organized for their use.

**Table 2.** Study participants and approaches.

| Aspect of Study | Approach |
| --- | --- |
| Process analysis for assessing needs | Participant-observations of meetings (N = 12), including informal discussions with meetings participants (between 4–28 participants in each meeting); Structured interviews (N = 8); Focus group discussions with community networks and QEF agents (N = 3) |
| Mapping of QEF and community knowledge processes, data, and concepts | Focus group discussions with community network & QEF domain & process experts (N = 4) |

Second, we developed formal maps of processes, data, and concepts, enabling us to specify relationships between healthy eating and physical activity behaviors, and related concepts that are central to the QEF mission, identifying which indicators measure each concept, for defining which data sources are required to calculate each indicator. Development of the conceptual maps entailed interviews and focus group discussions with QEF staff and community network members.

Future work will involve ontological analysis to capture the domain knowledge and specify relationships between actions and outcomes, providing semantic support for defining objectives, strategies, actions, indicators, and data sources, and facilitate interactive browsing by decision-makers in the MSP in the form of concepts, instances, relationships, and axioms, which will be encoded using OWL 2 Web Ontology Language [17]. Our ontology also facilitates the integration and interpretation of community data, and manages semantic interoperability between different knowledge sources. We note limitations in our study, for example in the small number of community networks that were sampled, given the time and resource intensive nature of such research, which presents constraints. However, such limitations can be addressed by further research that builds on what we have done to date.

## 5. Results and Discussion of Community Network Processes

Our assessment of the community groups' needs during their creation of epidemiological profiles was grounded in Pettigrew's 1997 framework for process analysis [30], which has five guiding assumptions that we use in organizing our findings (Table 3). These are represented in four themes from our observations, interviews, and focus group discussions. The framework also assumes that organizational and institutional processes and outcomes are context specific, as any given number of factors may interact in a specific way in a given context. Thus, findings regarding these community processes cannot be assumed to be the same for communities in other contexts.

Table 3. Adaptation of framework for our analysis of MSP processes in three communities [19].

| Aspect of Community Processes | Description |
|---|---|
| Embeddedness | MSP processes are embedded in multiple levels of inter-related structures |
| Interaction between human agency & structures | During community network processes, there are interactions between human agency (*i.e.*, participants in processes) and structures (*i.e.*, QEF guidelines for community network processes, data sources available, time allotted) |
| Temporal interconnectedness & Non-linearity | At each structural level, community network processes are temporally connected, in that past, present, and future processes are related to each other. The links between the multiple levels of structures, events that occur, and community network decision-makers' roles emerge over time, in a complex manner that linear explanations are unable to account for |
| Complex links between processes and outcomes | Community network processes are linked to outcomes at multiple levels in multiple domains, such as changes in children's food purchase, eating and physical activity behavior that may be linked with economic, health, and educational factors. Conceptual mapping enables decision-makers to more easily visualize the linkages between determinant of such behaviors, and outcomes such as health (e.g., obesity), educational (e.g., drop-out), and economic (e.g., cost burden) |

*5.1. Embeddedness of Processes*

Our key finding highlighted that given ambiguities about concepts, and disagreement among MSP decision-makers who are embedded in processes that span multiple levels and structures, there is a need for knowledge tools that facilitate real time collection, organization, synthesis, and analysis of data across levels, to enable stakeholder decision-making in a timely manner. We note that community network processes around healthy diets and physical activity are embedded in contexts where efforts are being made to combat childhood obesity at local, regional, provincial, federal, and global levels, as previously discussed [8], with potential contradictions arising at each level. Even at a given level, multiple embedded processes occur.

In our study, we focused on three local-level processes. First, at a fundamental level, stakeholders from diverse types of organizations congregate in multiple MSP meetings to engage in decision-making for developing their community epidemiological profiles, bringing diverse worldviews and perspectives that were evident in their divergent views on what healthy eating and physical activity issues, related to childhood obesity, should be prioritized. We identified a sub-process of five steps over the course of MSP meetings that constitute the development of epidemiological profiles: (i) accessing and selecting data sources and information deemed by community networks and QEF as pertinent to healthy eating and physical activity in the community; (ii) organizing the data and information (*i.e.*, mostly summaries or notes from the studies/reports reviewed); (iii) visualizing and analyzing the data and information; (iv) developing epidemiological profile documents; and (v) engaging stakeholders for validating epidemiological profile documents. Figure 1 below summarizes the sub-processes and needs identified from our analysis.

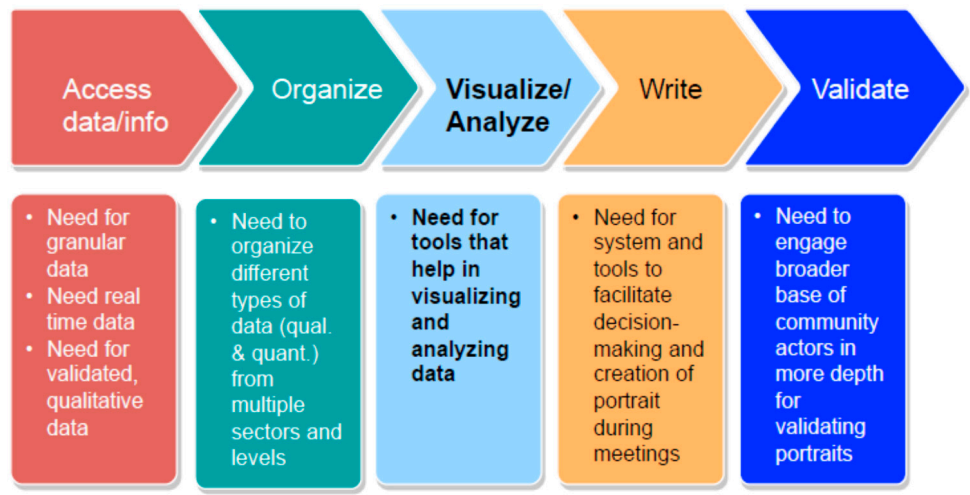

**Figure 1.** Community networks' sub-processes for developing epidemiological profiles.

Second, the various stakeholders also engaged in decision-making outside of MSP meetings, in their multiple roles in their respective organizations (See Table 4).

**Table 4.** Stakeholders and roles in QEF-funded community networks.

| Stakeholder | Description of Roles |
| --- | --- |
| Community network coordinators | Each community network has a coordinator, who is the key actor that incorporates data and information into community epidemiological profiles. |
| Community network members | They have multiple roles in their respective community networks, which vary greatly in size and organization, and include representatives from early childhood centers; primary and secondary schools; local, municipal, and regional education and health boards and agencies; community and voluntary organizations; municipal government; *etc*. Members have expertise in one or more of the following areas: healthy eating; physical activity; child development; urban development; transport; agrifood; planning and evaluation; communication; and issues of poverty |
| QEF development agents | Each agent typically works with about 5 community networks, and is embedded in a QEF regional office, which oversees the partnerships in a region (e.g., the QEF Montreal region office where our study drew its sample oversees 25 community networks) |
| QEF evaluation team | Provides data and information for community networks to use for developing epidemiological profiles |
| Lucie & André Chagnon Foundation | Canada's largest private family foundation, which is the provincial government's partner in developing QEF as funding mechanism. Has six members that serve on the board of QEF. |
| Quebec provincial government | The provincial government has six members that serve on the board of QEF. |

Study informants highlighted that the decisions that stakeholders made were often based on the perceptions that they had from their respective organizations, rather than from hard data. For example, the following quote summarizes a point that was consistently made across the three communities in our sample: *We make decisions based on local information. The problem is that a lot of the information is based on people's personal judgment* (Community network member).

At another level, we note that the process of developing epidemiological profiles, is itself embedded within other processes (Figure 2): it is one among a seven-step iterative process by which community networks are mobilized with QEF support. The other six steps include using the epidemiological profile to inform a diagnostic analysis of what are the priorities to be tackled in the community, which is captured in a diagnostic report; developing a vision of change; developing a 3-year strategic plan; developing annual action plans with interventions; implementing interventions; and evaluating interventions. The community mobilization processes are themselves embedded in other processes underway at regional, provincial, federal, and global levels [8]. Informants noted that the multiple levels in which their processes are embedded requires that they are guided overall by consistent questions.

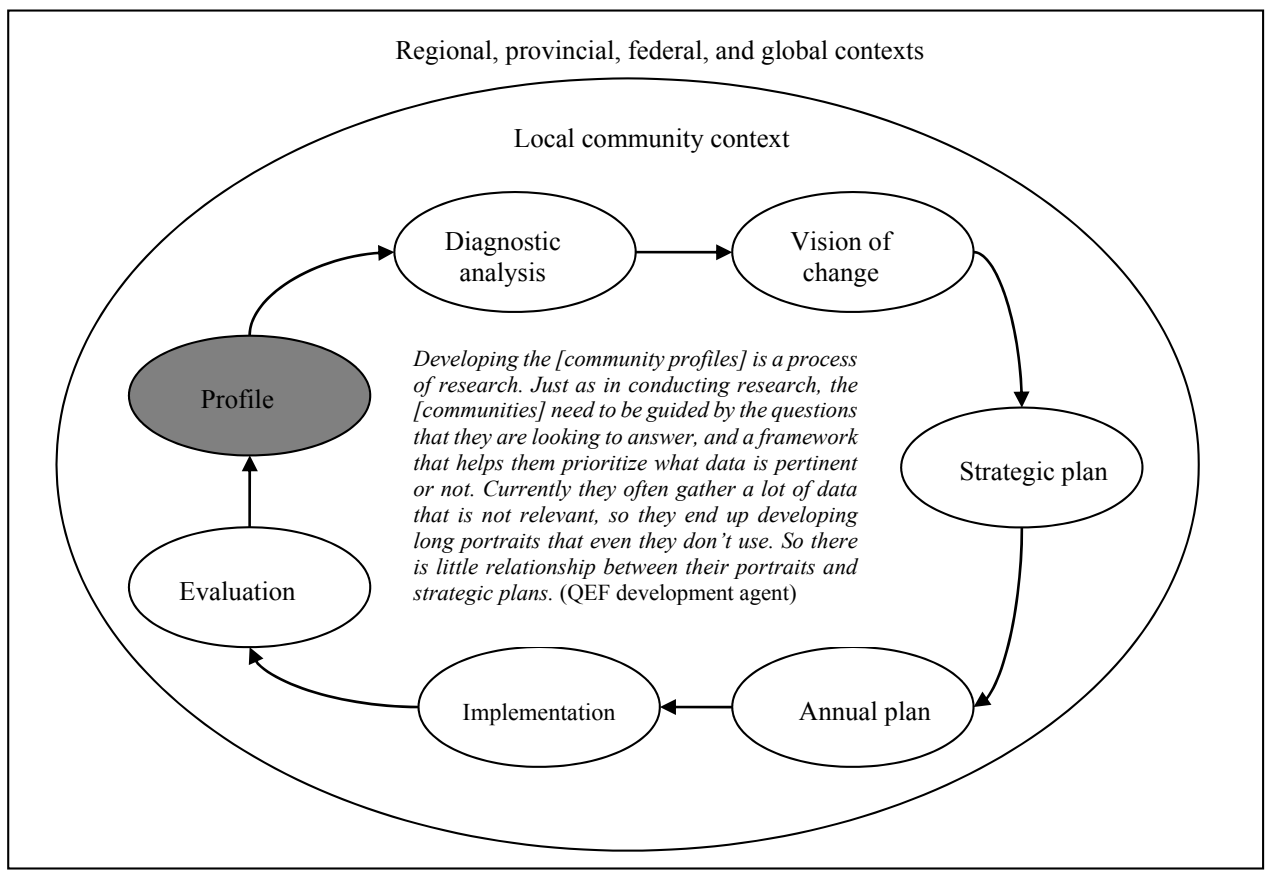

**Figure 2.** The embeddednes of community mobilization processes.

*5.2. Interactions between Structures and Human Agency*

Regarding the interactions between structures and agency of the actors noted above, accessing and selecting data sources and information within organizational boundaries is a key part of developing epidemiological profiles. Given their limited capacities to make sense of vast amounts of data, informants particularly highlighted the value of having readily available, organized data and information. QEF has provided five categories of data and information that the community networks can use in developing their profiles, with each category potentially informing decision-making at local, regional, and provincial levels (Table 5).

**Table 5.** The categories of data and information that QEF encourages community networks to gather.

| Categories | Types of Data |
|---|---|
| Category 1 | Behaviors and lifestyles of young people in healthy eating and physical activity |
| Category 2 | Characteristics of environments (*i.e.*, childcare, school, community services, municipal) that influence healthy eating and physical activity (political, economic, socio-cultural, physical) |
| Category 3 | Opportunities, levers and resistance faced in the community for influencing healthy eating and physical activity |
| Category 4 | Information on local stakeholders and their roles (organizations, consultative bodies, *etc.*) |
| Category 5 | Socio-demographic data, information and knowledge to compare contexts (deprivation, decay, cultural communities) |

In addition to the data noted above, community network members also seek and provide their own sources of data and information, which they may obtain within the structures of their own organizations. For example, in two of the community networks studied, some informants noted that their respective organizations developed their own epidemiological profiles for their domains of focus (*i.e.*, early childhood and consumption behavior around fruits and vegetables respectively), which, among other sources, then informed the overall epidemiological profiles of their respective community networks. Thus, the epidemiological profiles are potentially shaped through multiple interactions between the diverse actors and the structures within which they are embedded. Given the scope of this study, we focus on the data and information above that QEF encourages community networks to gather.

*5.3. Temporal Interconnectedness and Non-Linearity of Processes*

Regarding temporal interconnectedness and non-linearity in processes, we noted that the sub-processes that constitute epidemiological profile-development occur iteratively, over time. Prior to developing epidemiological profiles in a given cycle (e.g., 2014–2017), community networks conduct an evaluation of their activities over the previous funding cycle (e.g., 2011–2014). Thus, although our study was concerned with the development of epidemiological profiles for 2014–2017, we had to concern ourselves with processes that spanned two cycles: (i) the evaluation of interventions for the 2011–2014 cycle, (ii) the development of community epidemiological profile on healthy lifestyles for 2014–2017, and (iii) the diagnostic analysis that the epidemiological profile feeds into during the 2014–2017 period.

The timeframes and nonlinearity of community network processes was related to our finding that the mobilization of community members needs to be achieved in a more time-efficient manner. Notably, given capacity constraints of community network members, in developing epidemiological profiles they spend considerable amounts of time gathering data and information, which they sometimes subsequently do not use, with limited time and effort for mobilization around analysis of the data and information. To address this issue, informants noted a need for an information system that strikes a balance between being flexible, since there is great variability among the communities and their processes, and providing some structure by using data that has been previously gathered, organized, validated, and made easily accessible to help decision-makers (notably community network coordinators) to focus on analyzing information during the limited meeting times that they have to develop the epidemiological profiles.

*5.4. Complex Links between Processes and Outcomes*

Studying the development of epidemiological profiles, and community mobilization to prevent childhood obesity highlighted complex links between processes and outcomes at multiple levels, and across sectors. Thus, mapping processes and concepts enables decision-makers to more easily visualize linkages between determinant of physical activity and eating behaviors, and outcomes such as health (e.g., obesity), educational (e.g., school drop-out), and economic (e.g., cost burden due to obesity-related health problems, or drop-outs).

Informants—especially QEF development agents—noted that among community networks, a current lack of conceptual frameworks to guide decision-makers in their complex processes leads to disconnects between (i) evaluation of past interventions; (ii) the problems community networks choose to address, related to what data and information is accessed, selected and used during the development of epidemiological profiles, as well as (iii) the development of diagnostic documents, strategic plans, action plans, and interventions. QEF provides community networks with a framework that facilitates their epidemiological profile-development processes, by specifying the domains of issues and interventions, as well as the types of environmental/contextual factors (*i.e.*, political, economic, socio-cultural, and physical) for community networks to consider. Whereas this has been helpful, there is still a lack of conceptual clarity, given the diversity and complexity of multi-sectoral issues faced in communities.

To address the above issue, informants proposed that the processes of evaluating past interventions can be used for specifying questions and validating conceptual understanding of the relationships between interventions, determinants of healthy eating and physical activity, and desired outcomes in education, health, economic, and other domains, to inform what data and information community networks access, select, and use in the development of epidemiological profiles, diagnostic documents, strategic plans, and action plans.

Informants also noted that, at another level, the community networks are sometimes working with, and responding to the needs of multiple funders, and there is a lack of coherence in concepts and indicators that are being used in developing epidemiological profiles for such diverse funders. For example, speaking to a challenge that MSPs worldwide may face, a community network member noted that in their work on promoting fruits and vegetables, the indicators that they assess and report varies across donors, leading to more work than if there was coherence in how concepts and measures related to fruits and vegetables are perceived by QEF and other funders.

To address this issue, our findings suggested a process of concept mapping and ontology development by the various actors involved, including funders such as QEF and the community decision-makers. Such concept mapping and ontology development has to be done at multiple levels, first working in particular domains, and then across related domains. For example, developing conceptual mappings of healthy eating and physical activity interventions related to obesity prevention, as well as the outcomes associated, allows for assumptions and definitions about the relationships between concepts to be made explicit across the multiple sets of actors, and for specifying metrics to facilitate the work of funders, community groups, government, and all other concerned actors. We turn to this next.

## 6. Results and Discussion of Mappings

Based upon the community network needs identified from the process assessment presented in the previous section, input was requested from QEF and community networks in developing and validating mappings of processes, data, and concepts. Our assessment informed the development of "use cases" that were employed for the mappings that we discuss in this section.

*6.1. Process Mapping*

A process includes a series of actions performed to achieve a specific goal. A process map demonstrates the sequence of these actions. Process maps present an intuitive view for learning, analyzing and documenting the processes in a system. Findings from the first phase of the study informed our mapping of processes that we propose for making explicit how to manage data to effectively support communities in developing their epidemiological profiles (Figure 3). As illustrated in our mapping, we make explicit that processes involve two sets of actors: QEF and its higher level partners (*i.e.*, QEF and provincial government data management stakeholders (e.g., the provincial statistics agency, Institut de la Statistique du Québec—ISQ), domain and process experts, evaluation team) and the community networks (*i.e.*, community network coordinators and members, supported by QEF development agents). In "back-end" processes, QEF and its higher level partners engage in (i) accessing and selecting data during a first stage; and (ii) analyzing the data and making information available during a second stage; whereas (iii) the community networks on the "front-end" use the information that is made available for developing their epidemiological profiles.

The first set of processes, as demonstrated in Figure 3, begins with managers of an information system (a) accessing qualitative and quantitative data from multiple structured (e.g., databases) or unstructured (e.g., textual) sources. For example, there is data on eating habits from ISQ, the provincial statistics agency, including l'Enquête québécoise sur la santé des jeunes du secondaire, 2010–2011 (EQSJS); Enquête En Forme de Québec en Forme auprès des 10–17 ans, 2010–2014; data on schooling environment that is pertinent to eating, from the Bilan de l'implantation de la politique-cadre dans les écoles, of the Ministère de l'Éducation, du Loisir et du Sport (MELS); geographic data on municipal environment that is pertinent to eating, such as environment around schools (fast-food restaurants and convenience stores), from Géoclip, a data and information tool managed by the Institut national de santé publique du Québec (INSPQ); and food data, such as volumes of retail for different food categories, from Nielsen; *etc*. This is followed by the (b) normalizing of the data, after which relevant data is (c) selected, and (d) organized, with the output being a Data List.

In a second set of processes after the list of the organized data is created, (a) the data undergoes aggregation, followed by (b) verification. If the data is found not to be valid, it is communicated to experts for correction. Verified data is then (c) visualized and then (d) analyzed by experts, with the output being results of the analyses that are generated for creating a user-friendly application.

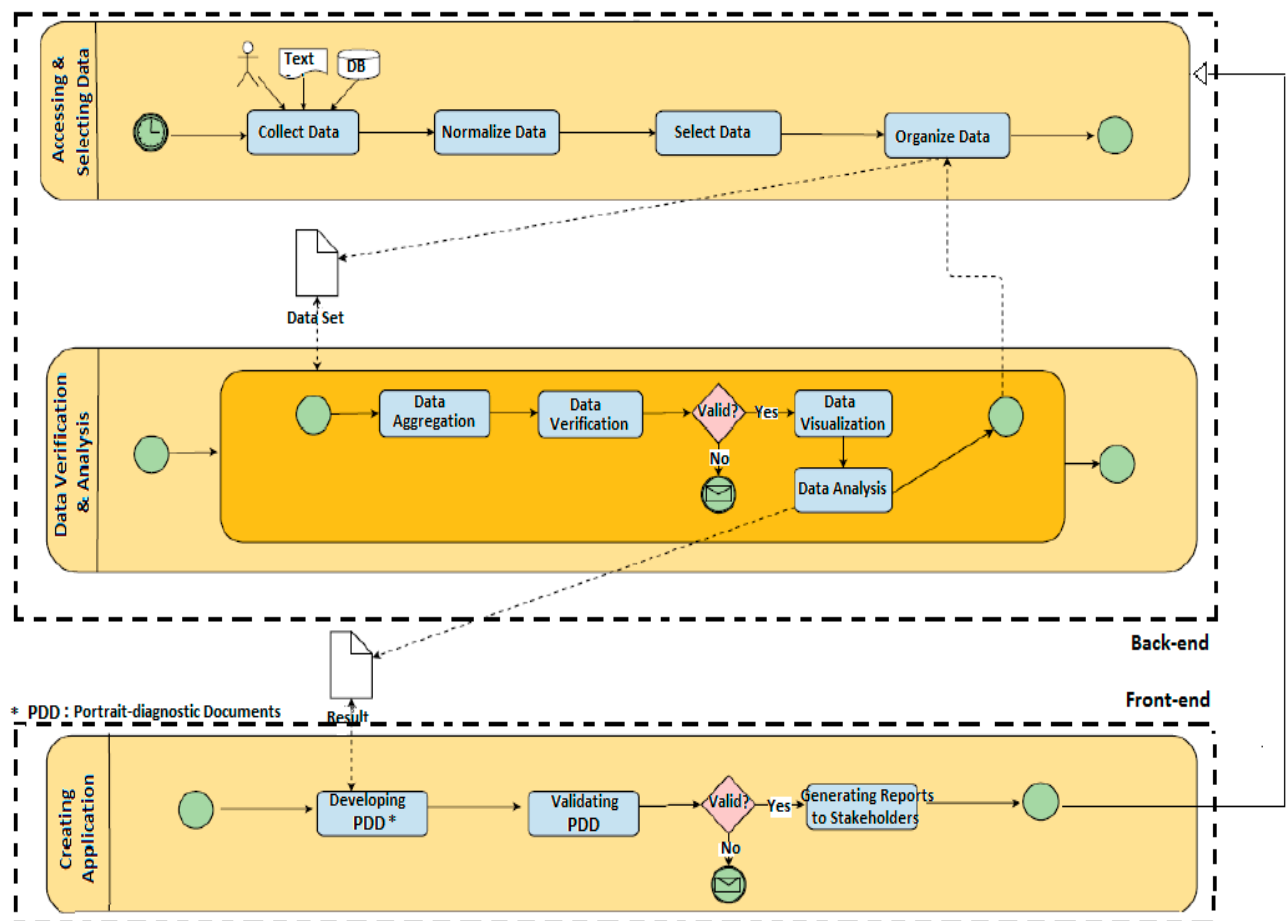

**Figure 3.** Mapping of observed processes for developing epidemiological profiles.

The third set of processes involves community networks use of the information generated from experts' analyses in (a) developing the epidemiological profile-diagnostic documents, which are (b) validated by the community network members. Where community networks indicate information is not valid, it is communicated to experts for correction. Validated information is then (c) used to generate reports for different stakeholders. The reports are also used by community network members in their evaluation processes.

Evaluation of data processes will be performed through using logical reasoners and expert reviews. To maintain the digital continuity of the proposed framework we follow the RLR method for change management and conflict resolution presented in prior projects conducted by members of this study team [31].

*6.2. Mapping of Data*

Currently the data used by community networks are available in different formats (e.g., textual, numerical, surveys and databases), and are distributed across various sources. In order to achieve an integrated view of the existing data sources they should be aligned consistently and mapped using the equivalent data elements. A data mapping highlights available data sources, and the links between them, which facilitates data integration by allowing a single interface for query and retrieval across the many disparate sources. Based on the information from QEF on what is available for community networks, the mapping below was developed (Figure 4). The five categories of information that QEF specifies from its community network evaluation framework are each linked to data sources that were further specified as needed.

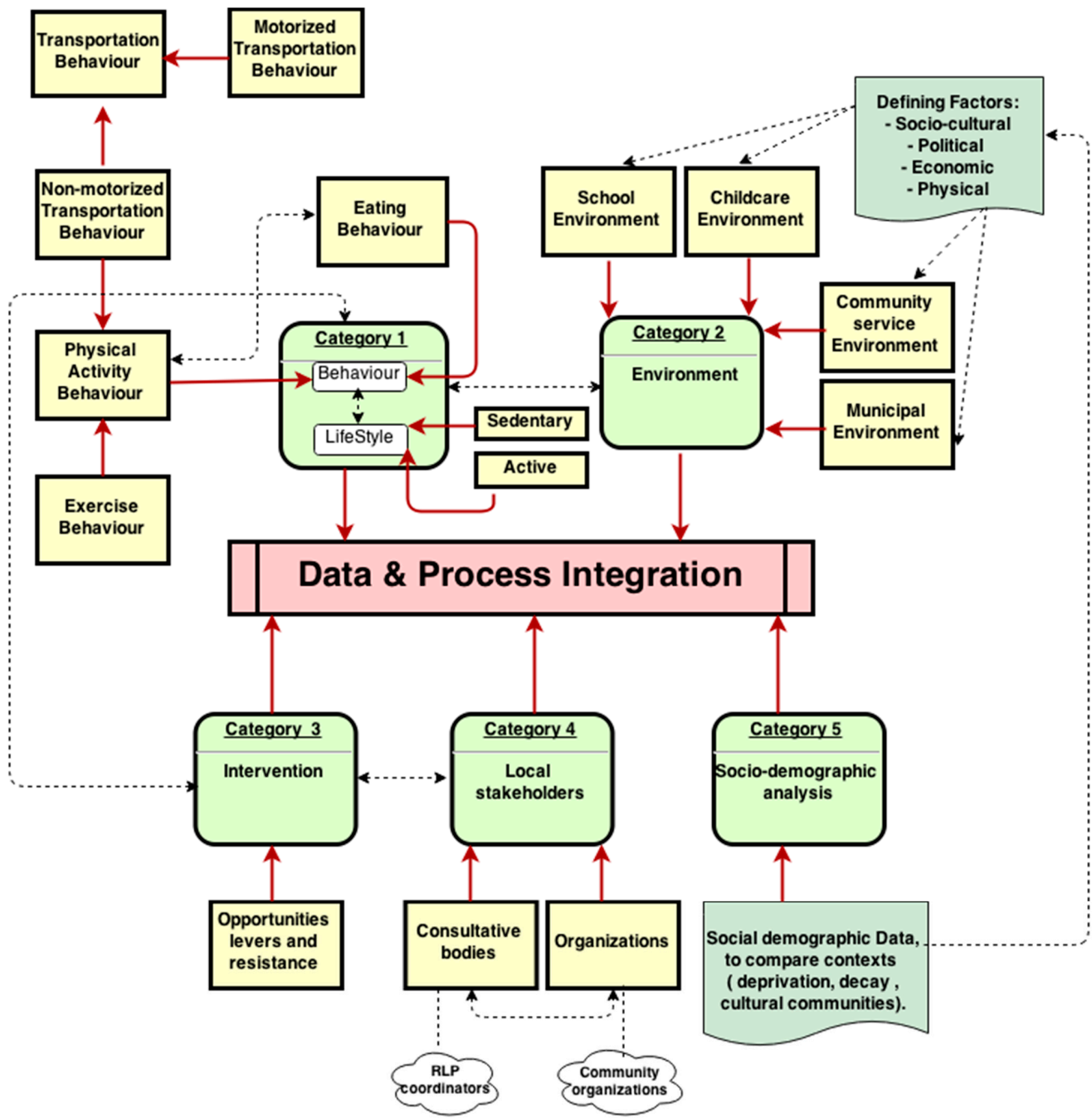

**Figure 4.** Data Integration—Mapping of QEF data for community networks (based on QEF Evaluation Framework).

For example, we noted that QEF's data on *behavior and lifestyles* (Category 1) should be specified as data on *behavior only*, as there are no data available on lifestyles, which is a different concept than behavior. Further, the data on behavior should be further filtered into data on *eating behavior* and *physical activity behavior*, and the latter can be further specified into *exercise behavior* and *transportation behavior*. *Environment* factors (Category 2) are related to *behavior* factors (Category 1), as illustrated in the dotted arrow between category 2 and 1. *Demographic* data on populations/neighborhoods (Category 5) are related to behavior factors (Category 1) as well as environment data (Category 2). We noted that whereas Category 5 data may be uniform in some communities (e.g., uniformly high levels of poverty), there is variation in other neighborhoods (e.g., socio-economic and educational backgrounds may vary).

There was disagreement among community networks about what constitutes Category 3 data, which was initially specified as "opportunities, levers, and resistance faced in the community", and with some informants expressing confusion about how that differs from environment data (Category 2). For example an informant noted that stakeholders in their community networks understood Category 3 data to mean that QEF is seeking data such as what other sources of funding there are in the community, which is information that they consider to be sensitive, and are thus not willing to share. We specified Category 3 as *Interventions*, noting that data on interventions/services offered by community networks, which is collected be explicitly categorized as such for Category 3 data. Linking data on interventions that is obtained from the community networks with the other types facilitates the collection and use of data in a manner that engages community networks.

There is also lack of clarity about what constitutes information on local stakeholders and their role (Category 4). Here also an informant noted that community networks understand Category 4 data to mean that QEF is seeking data such as which of the various stakeholders are responsible for specific issues being faced in the community, and what they contribute to the issues, which is information that they consider to be sensitive, and are thus not willing to share. However, informants noted that it is easier for them to map the various stakeholders around interventions in communities, as such a concept is more concrete, and they are able to more easily identify the various actors involved in interventions, as opposed to the more abstract task of mapping roles in the community generally.

*6.3. Conceptual Mapping*

Having mapped the processes and data used in developing the community profiles, we engaged in focus group discussions with QEF's domain experts to map relationships between concepts regarding healthy eating and physical activity, including the determinants of such behavior, which is related to childhood obesity. To do so, we posed queries that were collectively answered through concept mapping. For example: *What are the challenges related to the consumption of vegetables by children in our community, and how can community groups intervene?* Implicit in this broader question is the following more specific question, which can be answered through conceptual mapping: *What influences children to eat vegetables?*

Figures 5 and 6 below, present examples or concept maps, in different level of granularities, developed regarding the relationship between *Behavior*, in general, with other factors, as well as the interactions between different determinants of *Eating Behavior* with other parameters. We note that *Perceptions* (a concept) influence *Behaviors* (another concept), such as *Eating Behaviors*. Perceptions include *Perceptions of Social Norms*, which can be further specified into *Perceptions of Social Norms on eating behavior* (*i.e.*, perceptions of social norm among a population of children in which healthy eating is not "cool"). *Perceptions* themselves are associated with a *Population*, which has *Socio-cultural characteristics*. The *Built Environment*, including *Schooling, Childcare, Community Service*, and

Int. J. Environ. Res. Public Health 2015, 12

*Municipal Environments* also have an effect on *Behavior*. We also note that there are some relationships between types of behaviors, for example with *eating behaviour* being influenced by *purchasing behaviours*.

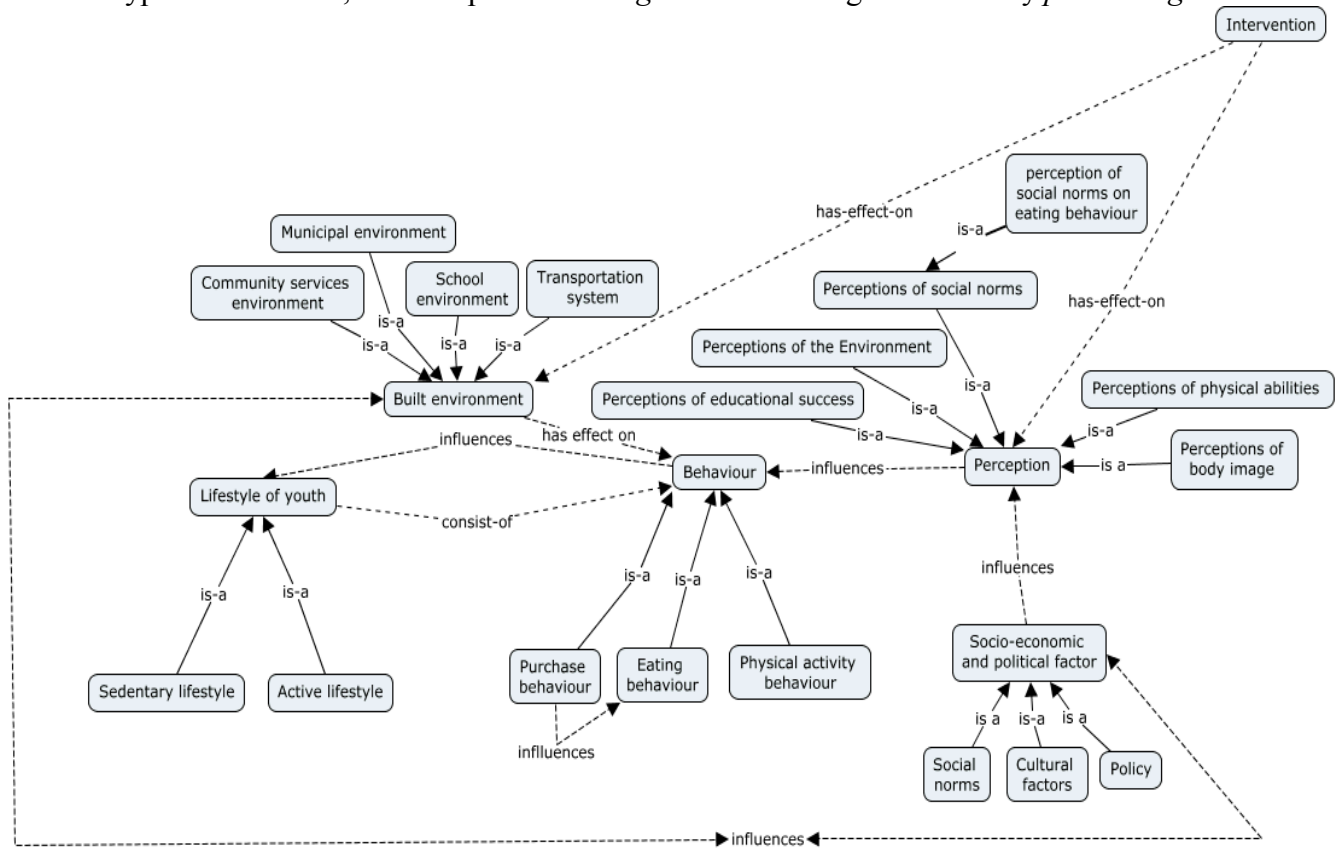

**Figure 5.** Conceptual Mapping (note that the solid lines represent subsumption (is-a) relationships, while dashed lines demonstrate associative relations (e.g., x *influences* y)).

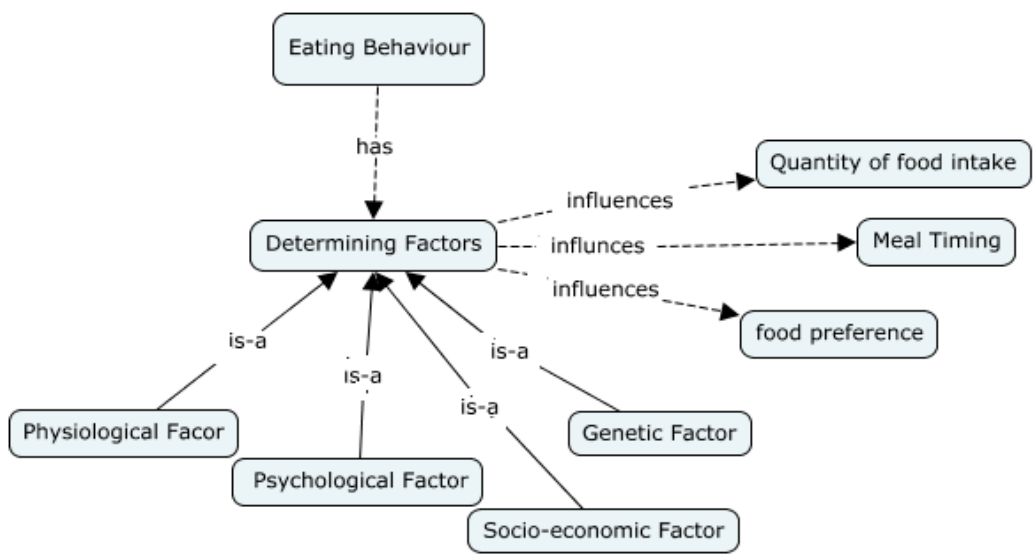

**Figure 6.** Conceptual map representing the interactions between determining factors of *Eating Behavior* with other parameters.

We note that *Behavior* in turn influences *Lifestyle of youth*, with all such linkages specified in conceptual mappings, which can be extended to mappings that include *Interventions*, and *Outcomes*.

## 7. Conclusions and Future Work

Childhood obesity is a global health crisis [32], and is also a risk factor for many chronic diseases such as diabetes, high blood pressure, stroke, heart disease, and cancers. Confronting this complex health issue requires multifaceted transdiciplinary programs. To be effective, these large programs must be based on a sound logical model of the different drivers, parameters and resistance in the domain. The process maps demonstrate a sequence of actions necessary to perform a specific process, namely developing a community portrait. The conceptual model presented in this paper provides an overview of the interactions between different components in the domain of childhood obesity. With such a semantic foundation, programs can implement interventions to control and reduce incidence, disability, cost, and mortalities globally. The major contribution of our research towards this goal is to propose a process model for capturing the knowledge and indicators needed to inform the multi-stakeholder partnership decision-making process at the community-level in Quebec, Canada. We expect that the knowledge will also be of use in other geographical settings with similar MSP decision processes. Future work will also focus on defining specifications for the development of a database of indicators and an information system to help decision-makers to view, analyze and organize indicators for their community. The results of this work should improve MSP decision-making to identify and implement effective interventions to address childhood obesity. We also plan to re-use some of the elements related to determinants of health and population health indicators currently implemented in the Montreal region Population Health Record (PopHR) [33], which computes and contextualizes indicator values using existing epidemiological knowledge. Using the PopHR framework enables us to analyses the causal relationships between different indicators, which is a crucial task for studying and evaluating different childhood obesity intervention outcomes.

## Acknowledgements

This research has been funded through Québec en Forme, the Fonds de recherche du Québec—Société et culture (FRQ-SC, Grant No. 2015-SE-179342), and the Social Sciences and Humanities Research Council (SSHRC, Grant No. 435-2014-1964).

## Author Contributions

Nii Antiaye Addy collected, reviewed, interpreted and curated data and drafted the manuscript, Arash Shaban-Nejad performed the computational data analysis, conceptual modeling and process mapping and contributed in writing the manuscript. Both Nii Antiaye Addy and Arash Shaban-Nejad contributed in the study design. David L. Buckeridge and Laurette Dubé conceived of the study, participated in its design and coordination and revised the manuscript. All authors read and approved the final manuscript.

**Conflicts of Interest**

The authors declare no conflict of interest.